\documentstyle[aps,prl,epsfig,manuscript]{revtex}
\begin{document}
{
\title{Measurement of the 1s-2s energy interval in muonium}
\author{
V. Meyer $^1$,
S.N. Bagayev$^5$,
P.E.G. Baird$^2$,
P. Bakule$^2$,
M.G. Boshier$^4$,
A. Breitr\"uck$^1$,
S.L. Cornish$^2$,
S. Dychkov$^5$,
G.H. Eaton$^3$,
A. Grossmann$^1$,
D. H\"ubl$^1$,
V.W. Hughes$^6$,
K. Jungmann$^1$,
I.C. Lane$^2$,
Yi-Wei Liu$^2$,
D. Lucas$^2$,
Y. Matyugin$^5$,
J. Merkel$^1$,
G. zu Putlitz$^1$,
I. Reinhard$^1$,
P.G.H. Sandars$^2$,
R. Santra$^1$,
P. Schmidt$^1$,
C.A. Scott$^3$,
W.T. To\-ner$^2$,
M. Towrie$^3$,
K. Tr\"ager$^1$,
L. Willmann$^1$
and
V. Yakhontov$^1$
          }
\address{
$^1$ Physikalisches Institut der Universit\"at
Heidelberg, D-69120 Heidelberg, Germany
\newline
$^2$ Physics Department, University of Oxford, Clarendon Laboratory, 
     Oxford OX1 3PU, United~Kingdom
\newline
$^3$ Rutherford Appleton Laboratory, Chilton, Didcot, Oxon OX11 0QX, 
     United~Kingdom
\newline 
$^4$ Physics Department, University of Sussex,  Brighton BN1 9QH, 
     United~Kingdom
\newline 
$^5$ Institute of Laser Physics, Novosibirsk 630090, Russia
\newline 
$^6$ Physics Department, Yale University, New Haven Ct., 06520, USA
\newline 
            }
\maketitle                                                        
\begin{abstract}
{
The 1s-2s interval has been measured 
in the muonium ({$\mu^+e^-$}) atom by Doppler-free two-photon laser spectroscopy. 
The frequency separation of the states was determined to be 2\,455\,528\,941.0(9.8)~MHz
in good agreement with quantum electrodynamics.
The muon-electron mass ratio can be extracted and is found to be  206.768 38(17).  
The result may be interpreted as measurement of the muon-electron charge
ratio as $-1- 1.1(2.1)\cdot 10^{-9}$.\\ 
%
}
\end{abstract}
%
%
}

%
%
%
We report a high precision laser measurement of the 1s-2s interval in muonium 
\cite{Maa_94,Jun_91,Chu_89}.
The hydrogen-like muonium atom (M=$\mu^+ e^-$) consists of 
leptons from two different particle generations, an antimuon ($\mu^+$) and
an electron ($e^-$) \cite{Hug_90}. The energy levels of the atom can be
calculated to the required accuracy within the framework of bound state quantum electrodynamics
(QED) \cite{Pac_98,Kar_99}. 
From high energy electron-positron scattering  experiments it is known that leptons have 
no internal structure  down to dimensions of $10^{-18}$~m and 
may therefore be
considered point-like objects \cite{Mar_90}. 
Unlike in the natural atoms or in hydrogen-like systems, which contain at least one  hadron,   
there is no complication in the theoretical description of M 
associated with the finite size and the internal structure of its constituent particles 
\cite{Bos_96,Hub_98}.
Con\-tributions arising in weak interaction from Z boson ex\-change, and from
strong interaction due to vacuum polarisation loops containing hadrons, are well understood.
Furthermore, 
precision measurements of 
electromagnetic transitions promise accurate values for fundamental constants. 
In addition, the M atom is an ideal system for testing standard theory \cite{Maa_94,Liu_99},
fundamental symmetries in physics and to search for yet unknown interactions~\cite{Wil_99}. 

The M formation  techniques available at present  favour the 
production of atoms in the 1s state.
For this reason
the  ground state hyperfine structure interval 
\cite{Liu_99}
and the 1s-2s interval $\Delta \nu_{{\rm 1s2s}}$ \cite{Maa_94} have been measured 
with highest accuracy. 
The muon lifetime of 2.20~$\mu$s
limits the natural line width of both transitions to 145~kHz 
(FWHM). 

The M 1s-2s transition offers a clean determination of the muon mass
which is required for the interpretation of experiments
on the muon magnetic anomaly \cite{Car_99} and for a determination of the weak interaction
Fermi coupling constant G$_F$ from upcoming measurements of the muon lifetime \cite{tmu_99}. 
Neglecting hyperfine structure, the theoretical value for the 1s-2s interval is
$\Delta \nu_{{\rm 1s2s}}$(theory) = 2\,455\,528\,935.4(1.4)~MHz \cite{Pac_98,Kar_99,Pac_99}
and is mainly determined by the well known Rydberg constant \cite{Wil_98}.
There is  a significant 1.187~THz (4800~ppm) contribution from the reduced mass of the system. 
The accuracy of the theoretical value of $\Delta \nu_{{\rm 1s-2s}}$(theory)  is  dominated
by the present knowledge of the muon mass to 120~ppb \cite{Liu_99}.
The QED corrections amount to 7056~MHz (2.9~ppm) and are known to 20~kHz (0.008~ppb)
\cite{Pac_99}. 
They differ from those in atomic hydrogen (H) and deuterium (D),
where they have been verified to very high accuracy, 
through muon mass dependent terms. Here the largest such effect is 18.1~MHz 
due to relativistic recoil \cite{Pac_98,Kar_99}.

The $1^2S_{1/2}$,F=1--$2^2S_{1/2}$,F=1 two-photon transition
in M was excited Doppler-free with two counter-pro\-pa\-ga\-ting laser beams.
The experiment was performed using the intense
pulsed muon source at the ISIS synchrotron
of the Rutherford Appleton Laboratory (RAL) in Chilton, UK. 
The accelerator operates at 50~Hz repetition rate
and provides in its DEVA beam area 3500 $\mu^+$ per pulse at p=26.5~MeV/c momentum
and $\Delta$p/p=10\% momentum bite.
Muonium atoms were formed by electron capture after stopping the muons close to the surface
of a target of SiO$_2$ powder (Cabosil M5, Cabot Company). About 80 atoms per pulse
left the  target 
through its surface. They have a thermal Maxwell Boltzmann velocity distribution 
with an  average velocity $\overline{v}$=7.4(1)~${\rm mm}/\mu{\rm s}$ corresponding to
296(10)~K \cite{Woo_88}. 
The M production was monitored continuously by detecting positrons
from the decay $\mu^+ \rightarrow e^+ \nu_e \overline{ \nu}_{\mu}$ in M atoms 
in vacuo with a telescope consisting of two proportional wire chambers 
each with two planes of orthogonally oriented wires \cite{Woo_88}.
%
%
%
The atoms interacted 8~mm above the target surface with pulsed laser light at 244~nm
wavelength. This beam was reflected 5 times back and forth between two mirrors
to enhance the overlap with the atom cloud. A third mirror directed the 
beam back onto itself. On average there were 1.5(5) M atoms in
the laser beam per pulse.

\begin{figure}[htb]
  \centering
  \epsfig{file=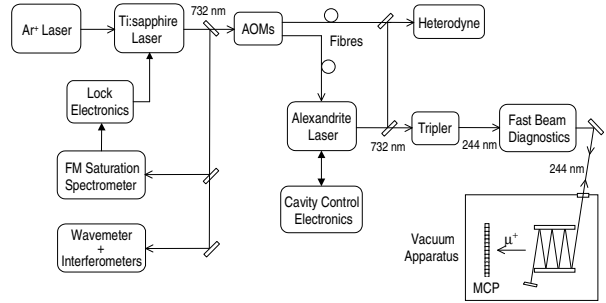,width=8.5cm}
  \caption{    Laser system. A Ti:sap\-phire laser is frequency offset locked to 
               a resonance in
               molecular $\rm{I}_2$. The cw light seeds an alexandrite ring laser 
               amplifier the output of which is frequency tripled using two LBO 
               and one  BBO crystals.}
  \label{m1s2s_laser}
  \end{figure}

The required UV light was generated by frequency tripling the output of a
 solid state pulsed laser system (Fig. \ref{m1s2s_laser}) .
The first stage is an Ar-ion pumped  cw Ti:sap\-phire laser (Microlase MBR-110).
At the fundamental wavelength of $\lambda_0$ = 732~nm 
it produced a beam with 0.5~MHz bandwidth and 500~mW output power.
The cw laser frequency was 
offset stabilised by double passing an
acousto-optic modulator (AOM) to a hyperfine  component of 
an R-branch line in $^{127}$I$_2$.
This was observed by Doppler-free frequency modulation spectroscopy
of thermally excited iodine vapour enclosed in a hot cell (650$^\circ$C) with a
cold finger at 39$^\circ$C. 
Suitable iodine reference lines were identified \cite{Lan_99} 
and their frequencies $\nu_{I_2}$ 
have been calibrated 
for M ($a_{(15)}$ component of the 5-13 R(26) line) and D
($a_{(19-21)}$ components of the 5-12 R(137) line) in preparation for this experiment
in collaboration with the 
National Physical Laboratory in Teddington, UK \cite{Cor_99}.

\begin{figure}[tb]
  \centering
  \epsfig{file=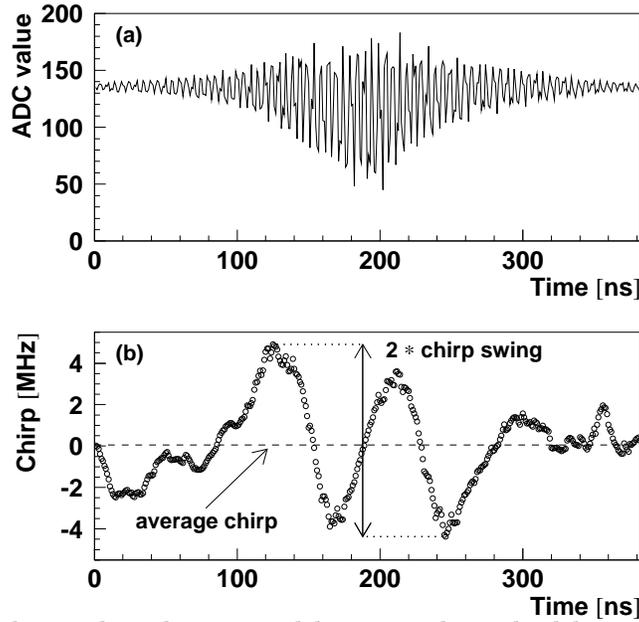,width=8.5cm}
  \caption{    A sample heterodyne beat signal between the pulsed laser output and a 
               frequency 
               shifted cw laser beam (a). The time derivative of its phase 
               yields the associated chirp, i.e. the shift of
               {\it instantaneous} light frequency (b). }
  \label{m1s2s_chirp}
  \end{figure}

In the following stage an alexandrite ring laser amplifier 
(Light Age PAL-101/PRO)
was seeded with 3.5~mW of cw light.
In order that the output pulse should coincide in time with the presence of M atoms,
the Q-switch was opened at a fixed point in the accelerator cycle, and a cavity length 
control system  was developed \cite{Bak_99} to ensure resonance with the cw seed at this time.
When the laser was operated at $\approx$1.2 times threshold inversion, seeded, 
single longitudinal mode, TEM$_{00}$, output pulses were produced on more than 95\% of shots,
typically of 35~mJ energy and 125~ns pulse length, at 25~Hz.
The heterodyne beat of the two lasers was recorded using fractions of each beam brought 
together on a fast photodiode (Hamamatsu S2381), with the cw frequency
shifted by 200~MHz. The high mode purity enabled clean records to be obtained, as shown
in Fig. \ref{m1s2s_chirp}. These signals, together with the time dependent intensity,
were digitised for every laser pulse, using 1~GHz (DL515, home made) VME flash ADC systems.
At the 732~nm wavelength of this experiment, changes in the refractive index
of the alexandrite rods during the pulse, due to inversion depletion \cite{Bak_99},
were large enough to produce a chirp swing of up to 120~MHz in the {\it 
instantaneous} frequency,
if not corrected. By applying fast high voltage ramps to a pair of electro-optic
modulators \cite{Rei_96} in the cavity the chirp swing  was reduced to $\approx$6.5~MHz,
with a 3.4~MHz wide distribution. The pulse to pulse distribution of the average chirp
was set close to zero with a spread of 7~MHz.

In the final stage, two 15~mm long lithium triborate (LBO) crystals were used to
frequency double with minimum walk-off the output of the alexandrite laser.
Mixing with the residual fundamental radiation in a 12~mm long    
$\beta$-barium borate (BBO) crystal
then gave the frequency tripled output at 244~nm. Pulse energy was 3.3~mJ and length 86~ns,
on average, and the spatial profile retained the smooth Gaussian character of 
the input.
After expansion in a telescope, the beam was sent some 10~m to the
interaction region, where its rms radius was 0.6(2)~mm. The profile of the
outgoing and the  retro reflected beams (Fig.\ref{m1s2s_laser}) 
were recorded on every shot using a  128$\times$128 pixel photodiode array
(EG\&G, Reticon RA0128N). 
The UV light frequency could be varied by setting the 
AOM frequency to any value
within a 60~MHz interval. 
%
%
%
%

The 1s-2s transition was detected through the photoionisation 
of the excited 2s state in the same 
laser field. The muon released in this process was electrostatically accelerated to 2~keV\
energy in a pulsed, two-stage device, which was turned on 50~ns after the end of every
laser pulse. When the laser was on, 
the residual electric field in the fiducial volume was kept below 5 V/cm
which limited a possible dc Stark effect to below 120~kHz.
The muons were guided along a 1.66 m long path
through a 90$^\circ$ electrostatic  mirror 
and  a 90$^\circ$ bending magnet, onto a microchannel plate detector (MCP).
The transport system was mass selective and had a 7\% momentum range. The MCP was
surrounded by scintillation detector telescopes covering 94\% solid angle for 
detecting the positrons 
from muon decays 
as part of an event signature. The muon detector was shielded by a 10~cm thick
lead enclosure and provided less than 2.8 background counts per day  in the
expected 74~ns wide time of flight window for the muons. This time interval corresponds to 
6 standard deviations of the muon time of flight distribution, which had an average 
of 1.16~$\mu$s.
The number of MCP counts as a 
function of laser frequency represents the 
atomic signal.
%
%
%
%

\begin{figure}[tb]
  \centering
  \epsfig{file=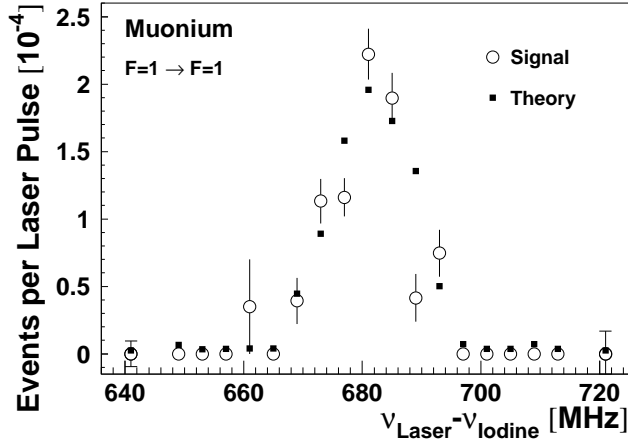,width=8.5cm}
  \caption{   Muonium 1s-2s signal.
              The frequency corresponds to the offset of  the laser at its fundamental 
              wavelength 732 nm from the 
              relevant reference in the $^{127}\rm{I}_2$ spectrum. 
              The open circles are the observed signal, the
              solid squares represent the theoretical expectation 
              based on measured laser beam parameters
              and a line shape model {\protect \cite{Yak_99}}. 
              }
  \label{m1s2s_signal}
  \end{figure}

For the M measurements 25 preselected values of the AOM frequency
were chosen. Every minute one of them was randomly chosen.
In total 3 million laser shots were fired
which produced a seeded pulse with a chirp swing below 15~MHz. 
Altogether 99 events were found  (Fig. \ref{m1s2s_signal}). 
To obtain the theoretical line shape, we calculated numerically for a 
randomly selected sample of 20\% of all the
laser pulses the probability for a resonant ionisation event
using a line shape theory  which
is based on a density matrix model. This allows the 
inclusion in each case of the recorded time dependent phase shift, the  intensity and 
the beam cross section for the laser light; details are given in reference \cite{Yak_96,Yak_99}.
We verified that the remaining 80\% and all the pulses which produced an event had the same
average distributions for chirp, chirp swing, intensity and spatial profiles. 
The signal amplitude was obtained from matching the integrals
over the experimentally observed and the theoretically predicted line shapes.
The line centre was obtained using the maximum likelihood fitting procedure
for a Poisson distribution \cite{Bra_76}
with the line centre as
the only free parameter. 
Data were recorded and analysed for two different M production targets 
within 34 hours of actual running. 
When both data sets were treated independently the line centres agreed within
3.7(9.5)~MHz; the statistical significance of both independent results can be 
verified in a Kolmogorov-Smirnov test which gives a 
significance level of 99\% and 44\%  respectively.
%
%
%

From the measured line centre frequency (Table \ref{m1s2s_table}) deductions of 976.4~MHz
for hyperfine structure and of
0.8~MHz for the second order systematic Doppler shift were made. 
The result is $\Delta \nu_{{\rm 1s2s}}$(expt.) = 2\,455\,528\,941.0(9.8)~MHz
for the centroid 1s-2s transition frequency. The contributions to the uncertainty are
9.1~MHz from statistics and 3.7~MHz due to randomly distributed systematic shifts
which include a 0.84~MHz calibration uncertainty of the I$_2$ lines, 
0.5~MHz due to frequency locking stability and a 3.4~MHz uncertainty caused by 
the residual linear Doppler shift 
due to the fi\-nite crossing angle of less than 55~$\mu$rad between 
the counter pro\-pa\-ga\-ting 
laser beams and 1.2 MHz from the estimated accuracy 
of the line shape calculations.

\begin{figure}[bt]
  \centering
  \epsfig{file=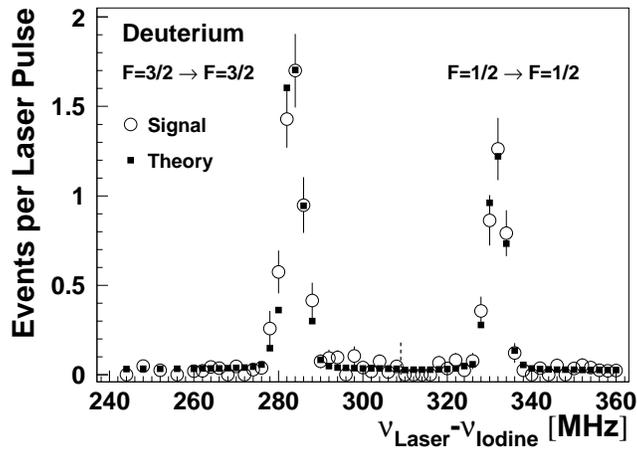,width=8.5cm}
  \caption{
               The experimental setup and the signal analysis procedures were tested and verified
               using signals from atomic D. The frequency corresponds 
               to the laser fundamental light at 729~nm and is the offset from the 
               relevant $^{127}\rm{I}_2$ reference line.
               Both hyperfine components were treated independently.
              }
  \label{m1s2s_deuterium}
  \end{figure}

The experimental setup has been tested and the analysis procedures were verified
using measurements of the two hyperfine components of the
1s-2s transition in D (Fig.\ref{m1s2s_deuterium}). The atoms were produced 
in a 99.99\% He and 0.01\% D$_2$ gas discharge and guided into the interaction region with a 
teflon tube. For this isotope, the density can be kept low enough at $<10^{-6}$~mbar residual 
gas pressure to avoid detector saturation and significant line shape distortion.
A transition frequency was obtained (see Table \ref{m1s2s_table})
in very good agreement with theory and previous
cw laser experiments. The dominant uncertainty arises from the calibration of the
relevant I$_2$ line.
The M--D isotope shift in the 1s-2s transition is 
$\Delta \nu_{1s2s}({\rm M-D})$ = 11\,203\,456.2(13.1)~MHz, where the uncertainty 
is to equal parts due to statistics in the M transition and the calibration of 
the iodine reference line for D.

Our experimental value of $\Delta \nu_{{\rm 1s2s}}$ 
agrees well with $\Delta \nu_{{\rm 1s2s}}$(theory) and with
earlier less accurate experiments \cite{Maa_94,Jun_91,Chu_89},
the accuracy of which had been affected strongly by laser amplifier 
chirp effects \cite{Rei_96}. 
The Lamb shift contribution to
 $\Delta \nu_{{\rm 1s2s}}$ 
has been extracted to $\Delta \nu_{{\rm LS}}$ = 7\,049.4(9.9)~MHz;
this is the most precise experimental Lamb shift value for M available today.
From a comparison between experimental and theoretical values for 
$\Delta \nu_{{\rm 1s2s}}$
we deduced the
muon-electron mass ratio as $m_{\mu^+}/m_{e^-}$ = 206.768\,38(17)
in agreement with $m_{\mu^+}/m_{e^-}$ = 206.768\,277(24) 
from the most recent determination of the muon magnetic moment
\cite{Liu_99}.

For hydrogen-like systems the leading order for the gross structure energy is proportional
to $(Z^2\alpha)\alpha/n^2$ where Z is the nuclear charge in units of the electron charge and 
$\alpha$ is the fine structure constant. By comparing $\Delta \nu_{{\rm 1s2s}}$(expt.)
and   $\Delta \nu_{{\rm 1s2s}}$(theory) we found for the $\mu^+$-$e^-$
charge ratio $Z= q_{\mu^+}/q_{e^-}=-1-1.1(2.1)\cdot 10^{-9}$. This 
is the best verification
of charge equality in the first two generations of particles. We note that the 
existence of one single fundamental quantised unit of charge is solely an experimental fact
for which no associated underlying symmetry has yet been revealed.

Our reported measurement here was statistics limited.  
In future the accuracy of the 1s-2s transition frequency could be improved 
using the novel technology employed here, with extended running 
and by a more restrictive selection for laser pulses with
low chirp swing. Significant progress could be expected from a cw laser 
experiment which  could be performed with higher numbers of M atoms produced by
high flux muon beams which may become available
at planned future accelerator sites such as a Japanese Hadron Facility, the
Oak Ridge Spallation Source and the front end of a muon collider.

%
%
%
We would like to acknowledge the support 
from the German BMBF, the EPSRC of the United Kingdom, the American NSF,
the European INTAS and NATO. The support from RAL staff members
and the very stable performance of the ISIS accelerator
were essential for the success of the experiment.
We thank K. Pachucki and S. Karshenboim for sharing with us their 
latest knowledge on the status of the theory and
J. Walling of Light Age Inc. and J. Hares of Kentech Ltd. for technical advice.

%
%

 %
\onecolumn
 \begin{table}[tbh]
 \squeezetable
 \caption{Contributions to the measurement of the 1s-2s frequency interval in
M and D and their uncertainties (all in MHz).}
 \begin{tabular}{|l||r|r|r||r|r|r|}
&\multicolumn{3}{c||}{muonium}&\multicolumn{3}{c|}{deuterium} \\
\hline
\hline 
& Frequency & stat. uncert. & system. uncert. & Frequency & stat. uncert. & system. uncert. 
\\
\hline 
$\Delta\nu_{measured}$ & $3\,113.6$ & $9.1$ & $0$ & $3\,144.5$ & $1.1$ & $0$ \\  
\hline
$+6\cdot\nu_{I_2}$ & $2\,455\,523\,890.2 $ & $0$ & $0.8$    & 
$2\,466\,730\,597.2$ & $0$ &$8.4$ \\
 \hline
$+$AOM and Lock Offset & $960.0 $ & $0$ & $0$ & $-1440.0 $ & $0$ & $0$ \\
\hline
+Stability of Lock & $0$ & $ 0 $ & $0.5$  & $0$ & $ 0 $ & $0.6$ \\
\hline
\hline
$=\nu_{1s-2s(F=1)}$ & $2\,455\,527\,963.8$ & $9.1$ & $1.0$ 
& $2\,466\,732\,301.7$ & $1.1$ & $8.4$\\
\hline
$+\Delta \nu_{hyperf.struct.}$& $976.4 $ & $0$ & $0$  & $95.5 $ & $0$ & $0$ \\
\hline 
$+\Delta \nu_{Quad.Doppler}$& $0.8$ & $0$&$0$ & $0.03$ & $0$&$0$  \\ 
\hline
$+\Delta \nu_{Res.Lin.Doppler}$& $0$& $0$ & $3.4$  & $0$& $0$ & $0.4$ \\
\hline
$+$ Uncert. Line. Calc. & $ 0$ & $ 0$ &$1.2$  & $ 0$ & $ 0$ &$1.2$ \\ 
\hline 
\hline
$=\nu_{exp}$& $2\,455\,528\,941.0$ &$9.1$ & $3.5$ & $2\,466\,732\,397.2$ &$1.1$ & $ 8.5$ \\ 
\hline
$\nu_{theo}$& $2\,455\,528\,935.4$ &$1.4$ & $0$ & $2\,466\,732\,407.7$ &$0.1$ & $ 0$ \\ 
\hline 
$\nu_{exp}-\nu_{theo}$& $5.6$ & $9.2$ & $3.5$ & $-10.5$ & $1.1$ & $8.5$\\
\end{tabular}
\label{m1s2s_table}
\end{table}

%
%
%
%
%
%

%
%
%

%
%
%
%

%
%

\end{document}